**Authors:**

*Askar Akaev*, Doctor of Sciences (Technology), Professor, aakaev@hse.ru. Institute of Mathematical Investigations of Complex Systems at Moscow State University and Laboratory for Monitoring of Sociopolitical Destabilization Risks of the National Research University Higher School of Economics (20, Myasnitskaya, Moscow, 101000, Russian Federation). The First President of the Kyrgyz Republic.

*Andrey Korotayev*, Ph.D., Doctor of Sciences (History), Professor, akorotayev@gmail.com. Laboratory for Monitoring of Sociopolitical Destabilization Risks of the National Research University Higher School of Economics (20, Myasnitskaya, Moscow, 101000, Russian Federation) and Institute of Oriental Studies of the Russian Academy of Sciences.

**Title:**

# GLOBAL ECONOMIC DYNAMICS OF THE FORTHCOMING YEARS. A FORECAST

**Abstract:**

The paper analyzes the current state of the world economy and offers a short-term forecast of its development. Our analysis of log-periodic oscillations in the DJIA dynamics suggests that in the second half of 2017 the United States and other more developed countries could experience a new recession, due to the third phase of the global financial crisis. The economies of developing countries will continue their slowdown due to lower prices of raw commodities and the increased pressure of dollar debt load. The bottom of the slowdown in global economic growth is likely to be achieved in 2017-2018. Then we expect the start of a new acceleration of global economic growth at the upswing phase of the 6th Kondratieff cycle (2018-2050). A speedy and steady withdrawal from the third phase of the global financial crisis requires cooperative action between developed and developing countries within G20 to stimulate global demand, world trade and a fair solution of the debt problem of developing countries.

**Introduction**

Starting from 2008, as a result of a change in Kondratieff wave[1] phases and shift in techno-economic paradigm (related to silicon semiconductor electronics), the world economy has been going through another cyclic crisis [*Glaz'yev, 1993, 2010*]. As forecasted [*Akaev, Pantin, Ayvazov, 2009*], the post-crisis depression having spread across the developed nations has been quite noticeable and would most likely last until 2017-2018. We expect an upturn in global growth to start in 2018 (see Fig. 1) on an upswing phase of the Sixth Kondratieff wave (2018-2040). We suppose that the innovative technologies of the sixth techno-economic paradigm will play a crucial role in ending the current financial and economic crisis and supporting further economic growth. NBIC technologies being the key technologies of the sixth wave of innovation [*Glaz'yev, 2010; Kazantsev et al., 2012; Koval'chuk, 2011; Rudskoy, 2007; Pride, Korotayev, 2008*] will become a sustainable source for economic growth and increasing competitiveness of the developed countries. According to another point of view, the key technologies would be MANBRIC-technologies which include medical, additive, information and cognitive technologies along with nanotech, biotechnologies and robotics (see e.g. [*Grinin, Korotayev, 2015a*]).

We witness to the implementation of Gerard Mensch's principle – 'innovations overcome the depression' – which was formulated during the global economic crisis of the 1970-s [*Mensch, 1979*]. Locomotives of the expansion phase of the sixth K-wave will be the USA, the European Union, Japan and Korea, the acknowledged leaders in terms of NBIC-technologies. At the same time, noteworthy is the fact that over the last years the center of global innovative activity has shifted to East Asia, where the number of patents granted annually is higher than in the rest of the world as a whole (see Fig. 1):

---

[1] For more detail on Kondratieff waves see e.g. [*Kondratyev, 1925, 2002; Akaev, 2010, 2011, 2013; Korotayev, Grinin, 2012; Korotayev, Tsirel, 2010; Korotayev, Zinkina, Bogevolnov, 2011; Korotayev, Grinin, 2012; Grinin, Korotayev, Tausch, 2016*].

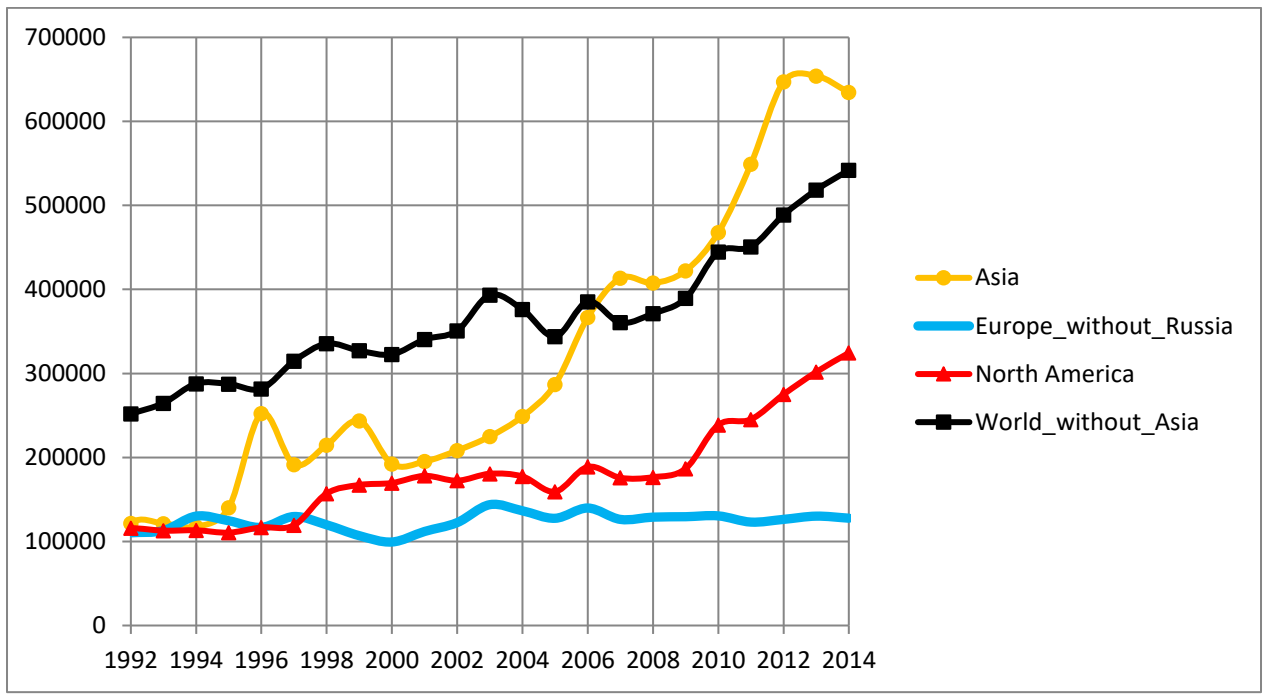

**Fig. 1. Number of patents granted annually in various world macro regions** (Source: [*WIPO, 2016*])

In 2014 Korea granted more patents for various inventions than entire Europe[2], whereas Japan granted twice as many patents as Korea. However, Japan and Korea are not the only ones that contribute to the leading positions of East Asia. China also contributes to those, having issued the same number of patents as Japan in 2014, which was two times more than that in Europe (see Fig. 2):

---

[2] Without Russia.

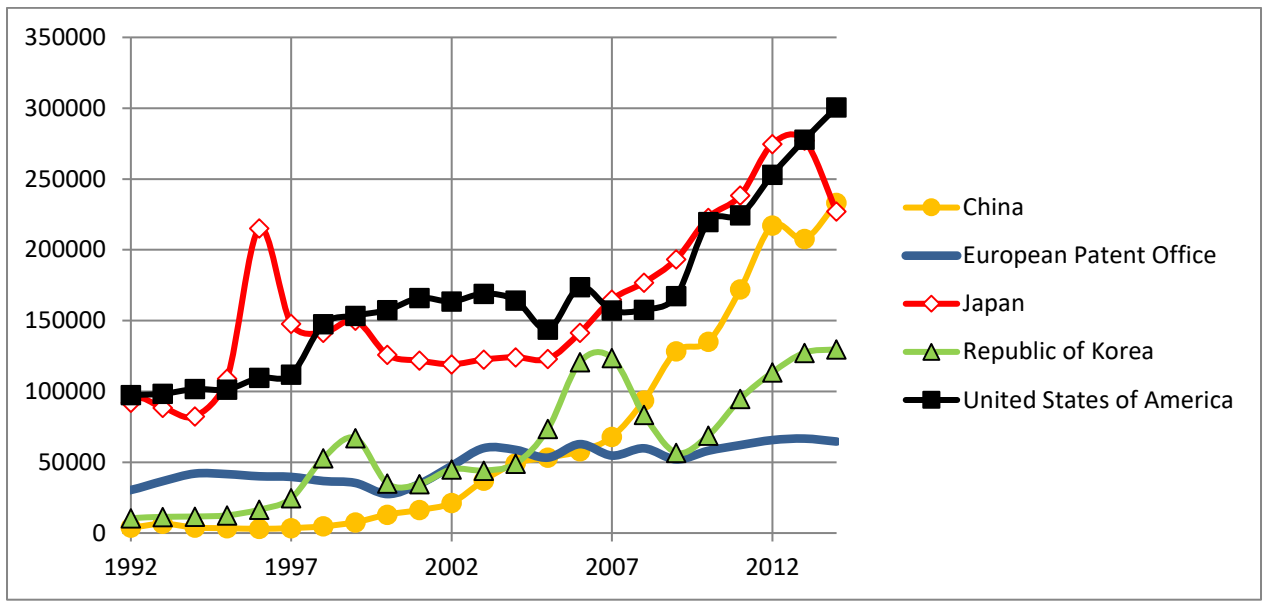

**Fig. 2. Number of patents granted annually in various countries, including European Patent Office** (Source [*WIPO, 2016*])

**Recession and depression of the fifth K-wave**

According to Akaev [*2013*], the fifth K-wave started in 1982[3]. Indeed, in 1982 global economic revival took place, which evolved into the longest (within the fifth K-wave) 12-year period (1982-1994) of sustainable and rapid growth at 3-4% annually. During that period investment was mainly made in fixed capital. World economy was truly prosperous between 1996 and 2006; productivity growth rates were twice higher than those of the period between 1985 and 1995. Starting from 2006-2007 growth rates declined among the OECD countries (see Fig. 4), which marked a transition from the upswing phase to the downswing phase of the fifth K-wave. Thus, 2006 became a climax of the fifth long cycle. Expansion phase of the latter lasted for 24 years. In just three years, the world faced the global financial crisis, which resembled that of 1929, which preceded the Great Depression of the 1930s, which justifies the name given to the recent crisis, *the Great Recession.* The crisis coincided with the depression phase of the Juglar cycle. According to experts, the upswing of the sixth K-wave is likely to start between 2018 and 2020.

---

[3] See [*Grinin, Korotayev, 2012*].

The beginning of the recession of the fifth K-wave, dated 2006, is confirmed by the dynamics of technical progress and investment. Thus, the USA total factor productivity, TFP, which is also called technological progress, began impeding in 2013 [*IMF, April 2015, p. 103*]. One should take into account, that the USA is considered to be world technological leader. In 2013 the impact of information and communications technology (ICT), which cannot be overestimated, started decreasing [*Perminov, 2007*]. By the mid 2000s an accelerated growth of TFP in ICT came to an end. Both production and capital influx to this sector reduced over the years preceding the financial crisis of 2007-2008 [*IMF, April 2015, p. 107*]. Hence, ICT being the technological basis of the fifth K-wave exhausted its possibilities to rise TFP, which marks the end of the upswing phase of the fifth long cycle and the beginning of search for new technological basis.

However, according to Akaev and Rudskoy [*2013*], ICT will help implement new key technologies of the sixth K-wave, being a drawbridge between the fifth and the sixth long cycles. It is crucial to support further development of ICT, as it will finally have a positive influence on TFP increase in the long run. Investments in fixed capital reduced after 2006, which marked a search for a new technological basis (see Fig. 3).

They decreased by 25% and have not reached initial level hitherto. However, after the boom of the 2000s in developing countries investment growth was not intermitted even at the times of crises (see Fig. 2), which is natural; K-waves come about earlier in developed countries as opposed to developing ones. A slight reduction in investment in developing countries after 2012 only partly reflects weak economic activity, having been a result of the financial crisis of 2008-2009. It is mostly a result of reducing prices for stock commodities, deterioration of financial environment, decreasing foreign direct investment (FDI), caused by the second wave of the world crisis after the fall of stock markets on August 4, 2011.

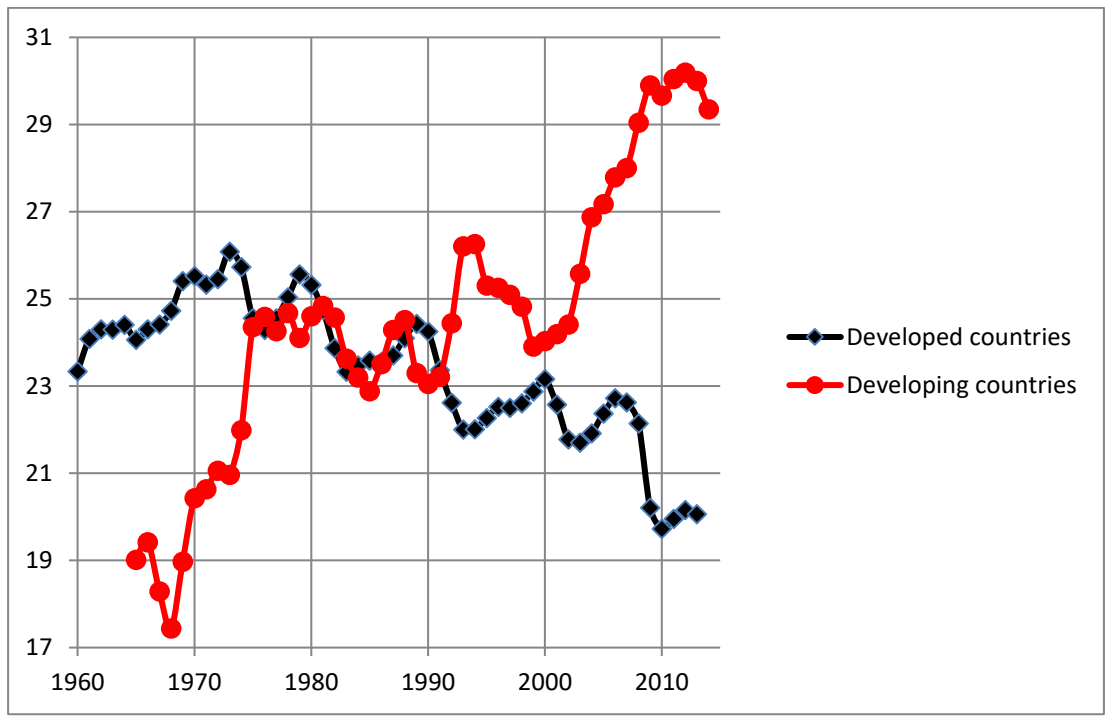

**Fig. 3. Dynamics of the gross fixed capital formation (as % of GDP) in developed and developing countries between 1990 and 2014** (Source [*World Bank, 2016*])

**Beginning of the second phase of the global financial crisis**

The date of the second wave of the global financial crisis, August 3, 2011, was forecasted by the authors in 2010, nine months before the outbreak of the crisis itself [*Akaev, Sadovnichy, Korotayev, 2010*]. In a series of works by the group of authors [*Akaev, Sadovnichy, Korotayev, 2010; Akaev, Fomin, Tsirel, Korotayev, 2010; Akaev, Sadovnichy, Korotayev, 2011; Akaev, Fomin, Korotayev, 2011; Akaev, Sadovnichy, Korotayev, 2012*] the forerunners of cyclic financial and economic crises were singled out; they included skyrocketing prices for oil and gold connected with creation and implosion of financial bubbles [*Akaev, Sadovnichy, Korotayev, 2010*]. It was further shown, that explosive price growth is well described by accelerating log periodic oscillations, coinciding with explosive growing trend, which is represented by a power-law function with quasi-singularity at a certain moment of time [*Akaev, Sadovnichy, Korotayev, 2011*]. Hence, one has to estimate the coordinates of the singularity point.

Causes and effects of the second wave of the 2011 financial crisis were investigated in our work [*Akaev, Korotayev, Fomin, 2012*]. It was shown that ultra-soft monetary policy of the USA and to a lower extent of other world powers resulted in an explosive growth of the world commodity prices. At times of crisis such policy was pursued to save financial and banking spheres. During the post-crisis period this policy stimulated economies. However, the cheap money flowed into stock and commodity exchanges and created "financial commodity bubbles". Their implosion in 2011 was associated with instability and economic downturn.

Despite quick and sustainable recovery after the global financial crisis of 2008-2009, developing countries demonstrated an impeded growth after 2011.

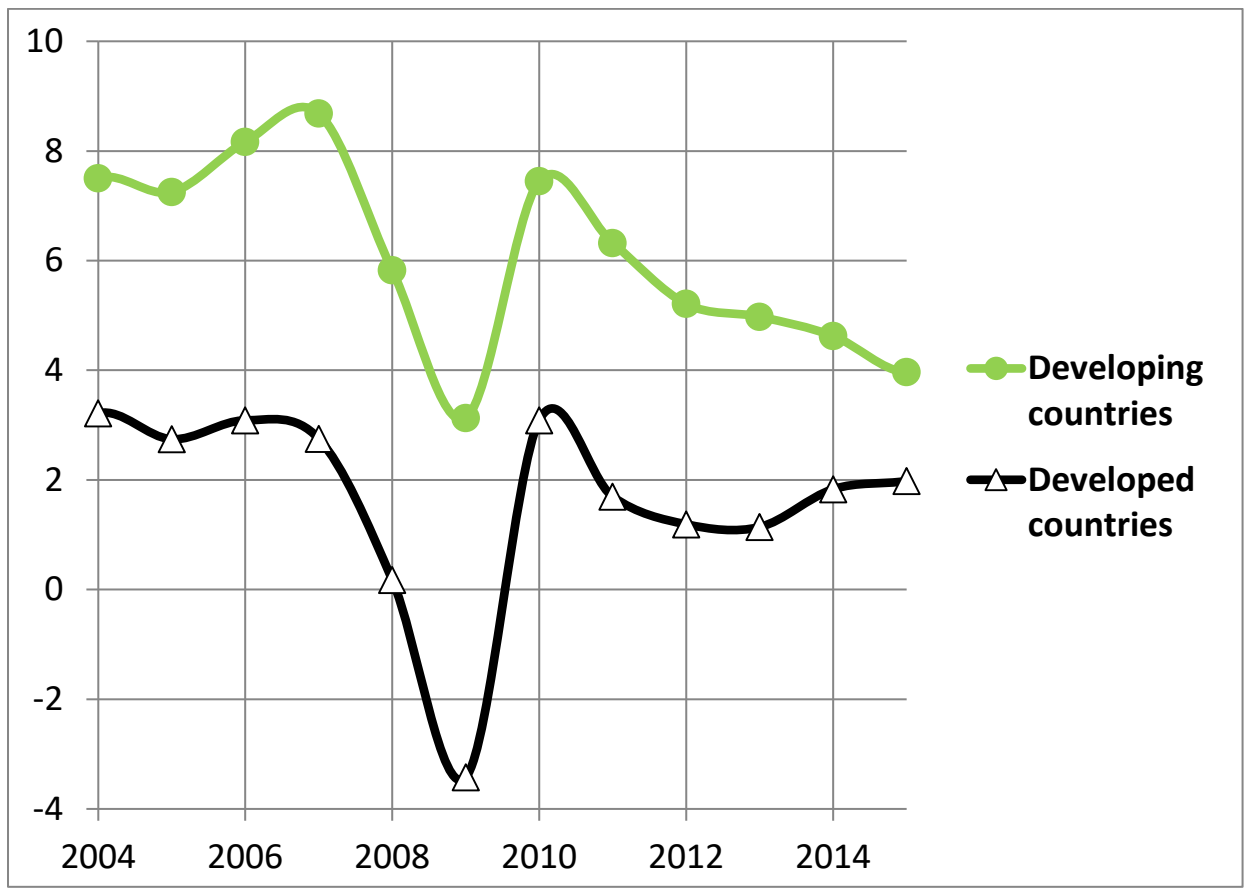

**Fig. 4. GDP growth dynamics of developed and developing countries between 2004 and 2015** (Source: [*IMF, 2016*])

In 2013 the average economic growth rates of developing countries were 1.5% lower than between 2010-2011, when the post-crisis recovery in the corresponding countries reached its climax. Moreover, this slowdown was almost simultaneous and affected almost every developing nation. Having started in the second half of 2011, the slowdown spread across 4/5 of the developing countries by the mid 2012. The first phase of the crisis resulted in falling standards of living in both developed and developing countries, which in its turn caused social upheaval in a number of developed nations; in various developing nations of the Middle East and North Africa social unrest turned so violent, that it practically became one of the causes of what later became known as "the Arab Spring" (see e.g. [*Korotayev, Zinkina, 2011; Korotayev et al., 2012; Grinin, Issaev, Korotayev, 2015*]).

On the verge of the second wave of the global crisis the world experienced a skyrocketing growth of prices for gold and other precious metals, food, energy carriers and other commodities. We have already mentioned the implosion of the "gold bubble" in August 2011. We identified, that implosions in various markets of stock goods came about simultaneously with bankruptcies and economic slowdown in developing countries, which are heavily dependent on trade in stock commodities. Apparently, such synchronous implosions and bankruptcies result in failures of established economic and trade chains. Thus, in 2011 that led to slump in world trade growth rates. For the first time within the last 30 years they stopped to exceed substantially world economic growth rates (see Table 1).

**Table 1**

**Dynamics of trade growth and economic growth in various countries and groups of countries between 2010 and 2015. IMF forecast up to 2020**

(%)

| Years/% | Actual growth | | | | | | IMF forecast | | | | |
|---|---|---|---|---|---|---|---|---|---|---|---|
| | 2010 | 2011 | 2012 | 2013 | 2014 | 2015 | 2016 | 2017 | 2018 | 2019 | 2020 |
| Growth rate of the world economy | 5,4 | 4,2 | 3,4 | 3,3 | 3,4 | 3,1 | 3,6 | 3,8 | 3,9 | 4,0 | 4,0 |
| Growth rate of the world trade | 12,5 | 6,7 | 2,9 | 3,3 | 3,3 | 3,2 | 4,1 | 4,6 | 4,6 | 4,7 | 4,6 |
| Growth rate of developed economies | 3,1 | 1,7 | 1,2 | 1,1 | 1,8 | 2,0 | 2,2 | 2,2 | 2,2 | 2,0 | 1,9 |
| Growth rate of developing economies | 7,5 | 6,3 | 5,2 | 5,0 | 4,6 | 4,0 | 4,5 | 4,9 | 5,1 | 5,2 | 5,3 |
| Growth rate of the US economy | 2,5 | 1,6 | 2,2 | 1,5 | 2,4 | 2,6 | 2,8 | 2,8 | 2,7 | 2,2 | 2,0 |
| Growth rate of German economy | 3,9 | 3,7 | 0,6 | 0,4 | 1,6 | 1,5 | 1,6 | 1,5 | 1,3 | 1,3 | 1,3 |
| Growth rate of the UK economy | 1,9 | 1,6 | 0,7 | 1,7 | 3,0 | 2,5 | 2,2 | 2,2 | 2,2 | 2,2 | 2,1 |
| Growth rate of Japanese economy | 4,7 | -0,5 | 1,7 | 1,6 | -0,1 | 0,6 | 1,0 | 0,4 | 0,7 | 0,9 | 0,7 |
| Growth rate of Chinese economy | 10,6 | 9,5 | 7,7 | 7,7 | 7,3 | 6,8 | 6,3 | 6,0 | 6,1 | 6,3 | 6,3 |
| Growth rate of Indian economy | 10,3 | 6,6 | 5,1 | 6,9 | 7,3 | 7,3 | 7,5 | 7,5 | 7,6 | 7,7 | 7,7 |

**Source**: World Economic Outlook URL: http://www.imf.org/external/pubs/ft/weo/2015/02/weodata/index.aspx (Access date: 11.27.16).

Indeed, over the last years, world trade growth rates have been significantly lower than economic growth rates, which has not been observed over the last 20 years. One of the main reasons for decreasing world trade growth rates were the consequences of the global crisis of 2008-2009. The crisis caused the drop in the world turnover. Between 1980 and 1993 during the upswing phase of the fifth K-wave world trade grew rapidly at 4.7% annual growth rate as opposed to 3% annual world GDP growth rate. The elasticity coefficient equaled to 1.6. Between 1994 and 2007 the global trade grew twice faster than the world production, which is validated by IMF data [*IMF, April 2015, p. 44*]. Between 1986 and 2000 world real GDP growth by 1% correlated with 2.2% world trade increase. That elasticity was significantly higher than that of the previous period between 1970 and 1985 and the following period of 2001-2013. The elasticity coefficient of those two periods accounted for 1.3. Current fall in turnover is the major threat for developing economies. Therefore, all the measures contributing to global trade growth will also contribute to overcoming the economic crisis.

**On the third phase of the global crisis and its causes**

As mentioned earlier, the developed economies including the USA will experience the third crisis phase, determined by mid-term economic downturn related to the Juglar cycle. The developing countries are also at risk of encountering the third phase of the crisis due to debt issues, which we mentioned in 2012 [*Akaev, Pantin, 2012*]. In October 2015 the Goldman Sachs think tank with the chief analyst of world stock markets Peter Oppenheimer as its head, analyzed economic slowdown in China and other developing nations and came to a conclusion, that the third wave of the economic crisis was on the way. They claim that the third crisis wave is a result of the first two waves, the US and the EU bank failures and the EU sovereign debt crisis. According to their position, the following scenario took place. In response to the first two debt crises the Central Banks, including the FRS, the ECB and others sharply decreased the discount rates. Investors profited from that and began funding developing countries, especially China. Most of the Chinese debts were cumulated after the crisis of 2008-2009, when Chinese authorities expanded lending to keep a rapidly growing economy buoyant against the background of the global financial crisis.

As a result, developing nations cumulated a huge dollar debt, which continues growing and becomes a concern on the world scale. Dollar debt in the form of dollar credits and obligations issued by the companies of developing nations has more than doubled over the last 5 years and totaled 3.3 trillion dollars by the beginning of 2016. At present the US dollar is regaining its strength, while investors dump their assets within developing nations. Thus, companies from developing countries experience difficulties servicing debts and finding funding for large projects of theirs. The problem of covering dollar debt by companies from developing countries escalated in 2014. Modest dollar strengthening caused fall of national currencies, which aggravated debt servicing. The companies started cutting their costs and laying employees off, which resulted in impeded GDP growth in developing countries. Hence, economic slowdown affected both developing nations and the world economy as a whole.

Total debt of governments, households, corporations and financial sector in developing nations increased by 1.6 trillion dollars in 2015 and reached 62 trillion dollars, exceeding their total GDP by 210%. Developed nations, on the contrary, cut total debt by 12 trillion dollars to 175 trillion US dollars. Such massive debt severely limits opportunities for borrowing to maintain further economic growth. According to IMF, by 2015 companies from developed countries cut their debt only by .4% to 87.4% of GDP, while companies from developing nations increased their debt by 6.7% up to 101.3% of GDP. Total corporate debt of the first 19 developing nations exceeded 25 trillion US dollars. Corporate debt of the least developed nations tripled after the crisis of 2008-2009 and accounted for 2.6 trillion US dollars. AS IMF points out, in 2016 and in the next three years all the borrowers will be obliged to repay their obligations debts and other credits. Between April and December of 2016 they had to pay 730 billion dollars, while the sum of 2017 debt totals over 900 billion.

Global economic and trade slowdown along with huge debts implies a new global crisis, the third after the Great Recession of 2009. Carmen Reinhart and Kenneth S. Rogoff claim, that after reaching a certain point the government debt impedes economic growth. According to their estimates, critical level of government debt is 90% of GDP. In this case, a substantial amount of budget expenditures is used to repay debts instead of investing in infrastructure and development. As we have pointed out earlier, vast majority of developing nations is already in danger zone, while the developed nations are approaching it. It is apparent, that debts are easier to repay, when the income grows. However, when debt reaches a critical point, economy, on the contrary, experiences lack of income. Hence, at present, we are at risk of mass defaults, which in their turn can pave the way for a new crisis.

Thus, heavy dependence of the world economy on credits is justly called "unhealthy" in the UNCTAD's 2015 report on trade and development. The report says, that if the world does not solve the "debt" problem, it will face a new sovereign debt crises, like the one Greece recently came through. According to UNCTAD data, the level of global debt rose 1.4 times from 142 trillion dollars in 2007 to 199 trillion dollars at the end of 2015, exceeding the world GDP accounting for 77.3 trillion dollars by 2.6 times. In the nearest future the world will experience the US dollar regaining its strength, low commodity prices and slow economic growth in both developed and developing countries. Under such conditions, the majority of companies will run into difficulties attracting funding to repay bank debts, which is likely cause corporate defaults in 2017. The latter would lead to further economic slowdown and financial instability.

All in all, current economic growth is not reaching its potential, which is caused by ongoing attempts to lower debt burden in the USA, the EU and the over-credited developing nations. Both developed and developing nations face falling demand, which influences world turnover. Economic slowdown is accompanied by falling household incomes, which might lead to social unrest. Most likely the world is now reaching the third phase of the world crisis after the first phase centered on the USA and the UK (2008-2009) and the second phase centered on the Eurozone and developing markets (2011-2012). The third phase will be mostly caused by huge debts of both developing and developed nations, excessive and ineffective investment and excessive production capacities, especially in China.

**Protracted strengthening of US dollar and low commodity prices**

One of the most powerful arguments is the strengthening of the US dollar that started in 2014. Moreover, after a period of cheap dollar (2000-2013) usually comes a period of strong dollar, which lasts for 8-10 years. Rising dollar and increasing US discount rate create a number of problems for developing countries. First, capital flows from developing nations to American market. Second, developing countries face difficulties serving rising dollar debt and borrowing to stimulate growth; discount rate growth will explode loan costs all over the world. Third, when dollar regains strength for a long period of time, commodities usually fall in price. Once investors are confident about strengthening of dollar, non-productive assets fall in price. Oil, gold and other commodities prove to be indicators of weak or strong dollar.

The US dollar is experiencing a long cycle of rising, which will last at least up to 2025. That causes a long cycle of low or moderate prices for commodities, which are the main assets of the majority of developing nations. Hence, low commodity prices will impede potential economic growth of developing countries in the mid-term. Real prices for quite a few commodities, especially metals, have decreased in comparison with its peak values of 2011. For example, prices for metals fell in price by more than 20% in 2015 and are expected to fall by approximately 10% in 2016. Food prices declined by 16% in 2015 and are expected to fall by 5% in 2016. While fall in oil prices is mainly connected with its overproduction, declining prices for metals are mostly correlated with a new economic development model of China [*Grinin, Tsirel, Korotayev, 2015*]. Indeed, Chinese share in consumption of metals has accounted for more than 50% of the world consumption over the last years. Hence, economic slowdown in China plays a crucial role in decreasing prices for metals. And this trend will remain for a long time.

So, the expansion phase (1998-2001) of commodity supercycle came to an end. Almost the entire commodity complex, especially metals, is imbalanced. The widely tracked S&PGSCI slumped by 13.6% in July 2015 and reached the minimum of 2002. Previous bottom was tracked in 1998 with preceding 30% slump. At the end of 2015 IMF revised its forecast of world economic growth in 2016 due to falling commodity prices and made it more pessimistic [*IMF, October 2015*]. IMF also pointed out, that the period of low commodity prices would be protracted, and the entire world would have to adapt to that. According to IMF, commodity exporters were going to face the worst prospects. In accordance with IMF estimates, commodity exporters would suffer from decreasing economic growth rates, which would account for 1% annually. Exporters of energy carriers were going to experience an annual economic slowdown of 2.25% [*IMF, October 2015*]. Thus, the slowdown is not only cyclical, but also structural. Therefore, IMF suggests that economic policy should be aimed at raising investment in innovations, in other words, raising productivity rates.

Kondratieff's theory describes cyclical changes of high and low prices for commodities [*Akaev, Sadovnichy, Korotayev, 2010; Grinin, Korotayev, 2014*]. Marchetti and Nakicenovic drew attention to periodic fluctuations of prices for main energy carriers, which coincide with the K-waves [*Marchetti, Nakicenovic, 1979*]. Such fluctuations usually last for 10-12 years and mark a structural change in energy consumption. We showed [*Akaev, Sadovnichy, Korotayev, 2010*], that these fluctuations signify world cyclic economic and financial crises. Indeed, when world economy experiences the upswing phase of a long cycle, world market conditions improve, while oil prices, in accordance with Kondratieff wave dynamics, remain at a low stable level, defined by the cost of production and transportation. Once market conditions dramatically worsen, downswing begins and capital flows into oil and gold, booming the prices for these goods (see Figs. 5 and 6).

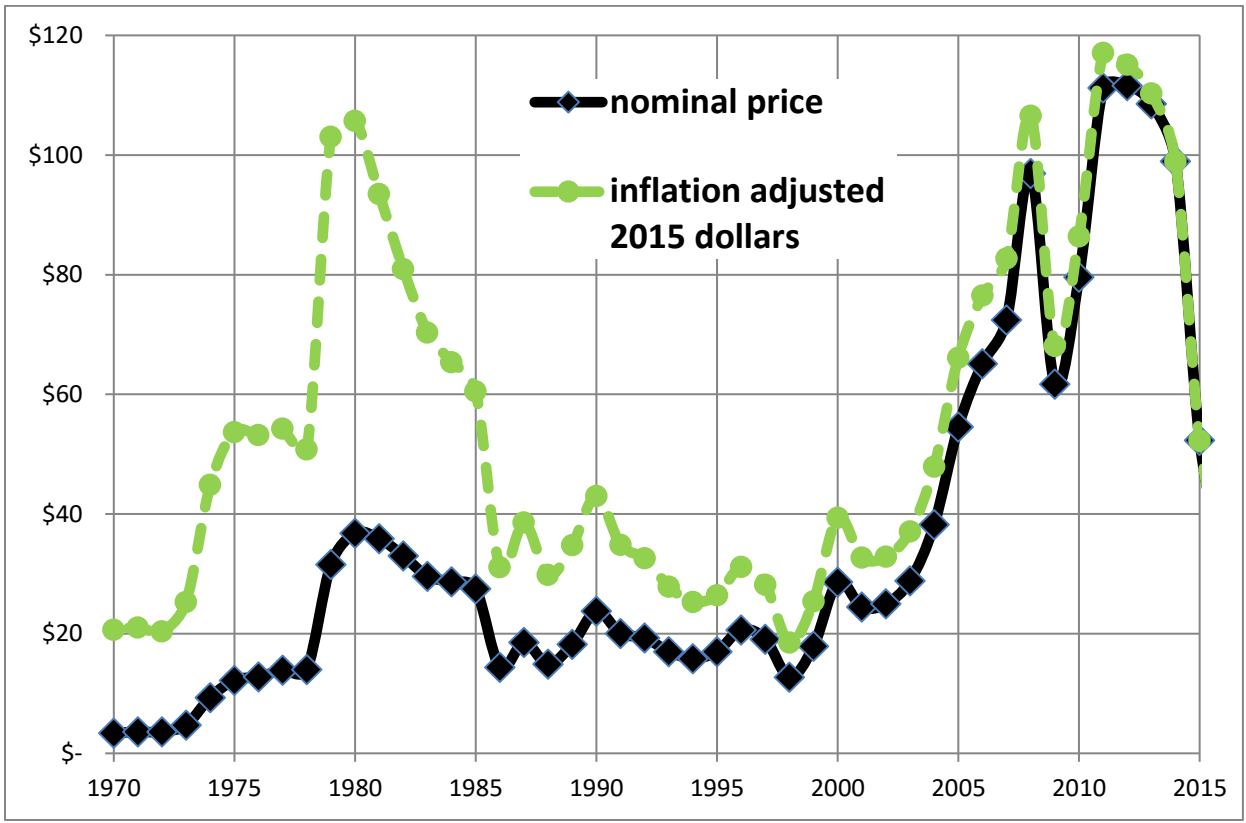

**Fig. 5. Dynamics of average annual oil prices between 1970 and 2015**
(solid line – nominal price; dotted line – oil prices in inflation adjusted 2015 dollars) *

*Sources*: database of Earth Policy Institute (Washington, DC, 2016). URL: www.earth-policy.org/datacenter/xls/update67_5.xls (oil prices between 1970 and 2006); database of U.S. Energy Information Administration. URL: http://www.eia.doe.gov/dnav/pet/pet_pri_spt_s1_a.htm (oil prices between 2007 and 2015); World Development Indicators Online (Washington, DC: World Bank, 2016), URL: http://data.worldbank.org/data-catalog/world-development-indicators (data on the US inflation).

---

* For the period between 1970 and 1973 the figure displays prices for light crude oil of Saudi Arabia. Between 1974 and 1985 it depicts purchase prices of American oil refineries for imported crude oil. Between 1986 and 2015 it describes average annual prices for Brent oil.

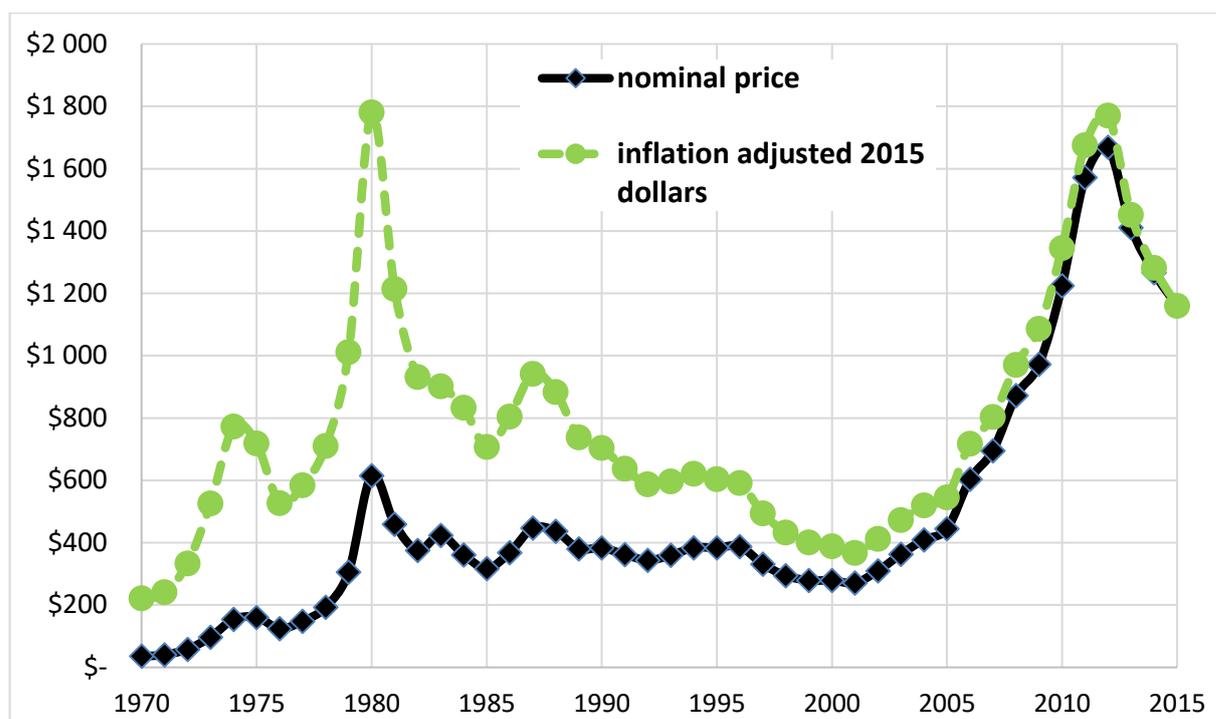

**Picture 6. Dynamics of average annual gold prices between 1970 and 2015, in US dollars**
(solid line – nominal price; dotted line – gold price in inflation adjusted 2015 dollars)[**]

*Sources*: database of the World Gold Council. URL: http://www.research.gold.org/prices/ (gold prices between 1970 and 2009); database of the USA Gold Reference Library. URL: http://www.usagold.com/reference/prices/history.html; World Development Indicators Online (Washington, DC: World Bank, 2016), URL: http://data.worldbank.org/data-catalog/world-development-indicators (data on the US inflation).

It is clear from Fig. 5, that after the oil crisis of the early 1970s, oil prices rose to $50 per barrel, later skyrocketed up to $95 per barrel in 1979, marking the beginning of the economic crisis of 1980-1982. Later, oil prices decreased gradually till the mid 1980-s, plunged between 1985 and 1986, then stabilized between $20 and $40 per barrel till 2003 and later began growing again. As we have mentioned earlier, between 2006 and 2007 the world economy faced the start of a downswing phase of the fifth K-wave, which is likely to last till 2017-2018. Deteriorating world market conditions accelerated oil and gold prices growth (see Fig. 10-11). From 2006 to July 11, 2008 oil prices grew rapidly from $60 to $147 per barrel (see Fig. 10). That was followed by the global economic crisis, which is usually assumed to have begun on September 15, 2008. Price for one barrel of Brent oil plunged to $40, then reached $70-80 per barrel, keeping this value through 2009, and later increased up to $100-120, stabilizing within this corridor between 2011 and 2013. Finally, in the second half of 2014 after the end of the program of quantitative easing and in anticipation of rising discount rate by FRS, oil started falling in price and reached $50 per barrel by the beginning of 2015. Average annual Brent oil price accounted for approximately $52 in 2015. However, in January 2016 oil prices slumped to $30 oil per barrel, reaching pre-2004 levels. Then the prices totaled $45 per barrel between March and May of 2016, which has been followed by a certain recovery to over $50 in direct connection with the activities of both OPEC and non-OPEC oil exporting countries.

---

[**] Average annual prices at the London Exchange.

Nevertheless, oil market supply exceeds demand, which is why the majority of forecasts for 2016 predicted the prices of $45-55 per barrel, which mostly has turned out to be correct. Hence, 12 years have passed since significant increase of oil prices. Moreover, that period almost entirely coincided with the recession phase of the fifth K-wave (2006-2018). It is logical to expect moderate oil prices and prices for energy carriers and other commodities including gold, other precious metals, food etc. at the beginning of the expansion phase of the sixth K-wave (2018-2040). Under favorable conditions, this period can last up to the 2030s. Moreover, oil prices are most likely to stabilize between $40 and $80 per barrel and to reach the mentioned ceiling by the end of the period. Annual inflation of the US dollar is supposed to account for 2%. It is also taken into account, that oil price corridor during the expansion phase of the fifth K-wave (1982-2006) was between $20 and $40 per barrel (see Fig. 10).

Decreasing prices for commodities caused falling national currencies and economic slowdown in developing countries. "Commodity currencies" are heavily dependent on raw material prices, so devaluation of national currencies in developing countries is a natural reaction to decreasing world prices for their main export goods, which allows avoiding significant production fall and unemployment growth. That is the way developing nations follow to adapt to new economic conditions. However, significant decrease of national currencies exchange rates entails a risk of negative outcome for enterprises' budgets, which hampers dollar debt repayment.

The only thing left is to define the beginning of the third phase of the financial crises. In order to do so, we should investigate a short-term prospect of the US economic growth.

The US economy has demonstrated relative stability over the last years. At the beginning of 2016 IMF forecasted the US annual economic growth of 2.6% in 2016 and 2017, which would surpass 2015 2.5% [*IMF, January 2016, p.7*]. Three factors contributed to such a positive forecast, which were rising demand, based on strong consumer expectations, rising employment rates and growth of the real estate market. In January 2016 unemployment level declined to 4.9%, reaching the minimum over the last decade. However, this was accompanied by rising concerns about potential interruptions in the US economic growth, when the FRS first after the crisis of 2008-2009 raised the discount rate on December 12, 2015. Experts even predicted recession. They supposed that the US economy might not stand global economic downturn, if the latter continues in 2017. Further downturn of, say, Chinese economy can lead to economic slowdown of American key partners, which are Canada and Mexico.

At the same time, there are several inner risks that include instability of the stock markets in the first place. Indeed, American stock market is on the verge of crisis. At the beginning of 2016 American stock indices showed the largest fall over the last 85 years. If the FRS raises the discount rate in 2017, stock markets are most likely to fall. In this case, low oil prices will lead to significant decrease of investment in energy sector, while strengthening dollar will limit American export. Moreover, annual inflation rate has remained much lower than an optimum level of 2%. According to the FRS forecast, in 2016 basic inflation will account for 1-1.5% and reach the target of 2% only in 2017.

Taking into account all the aforementioned risks, in March 2016 the FRS changed its forecast of the USA economic growth rate, decreasing the latter to 2.1-2.3% instead former 2.3-2.5% predicted in December 2015. Recently, OECD also changed its economic forecast of the US growth rates, lowering them from 2.5% to 2%. As is known, the lower the trajectory of US economic growth, the more vulnerable it is to external shocks. We also suppose that US stock markets will experience a fall in 2017. Due to further slowdown the US economy will face an economic downturn in accordance with mid-term Juglar cycles (7-11 years).

We will try to calculate the potential date of this crisis on the basis of Dow Jones Industrial Average (DJIA). Starting from the spring of 2009, simultaneously with the policy of quantitative easing, a new stock market bubble was created (see Pic. 7).

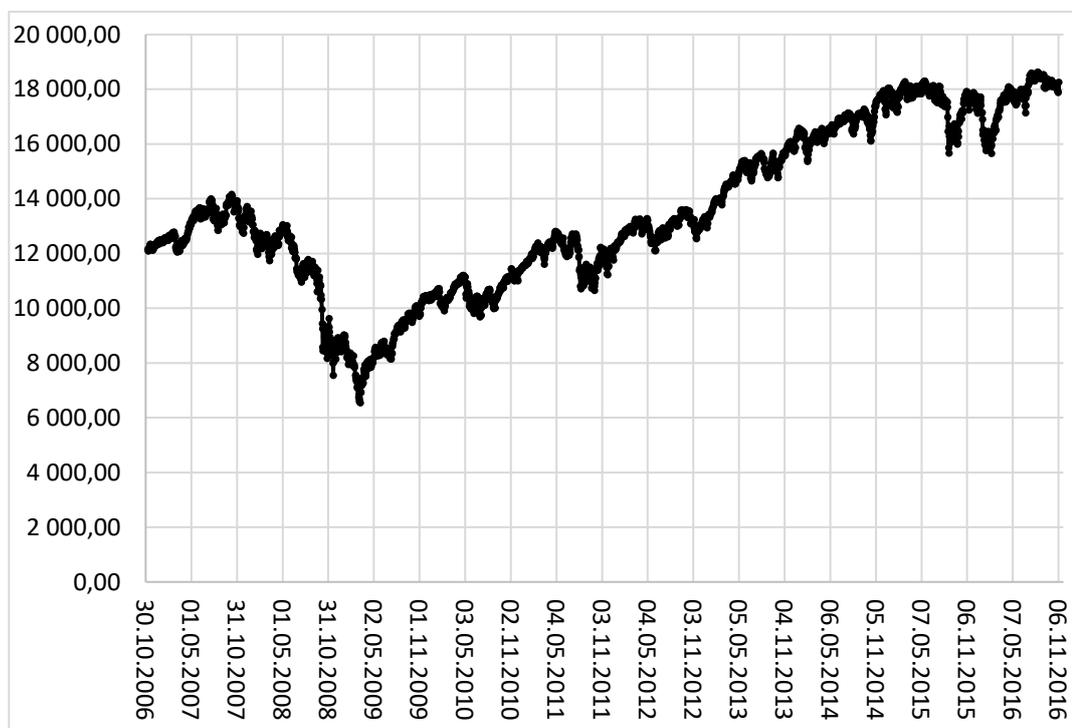

**Fig. 7. DJIA dynamics between fall of 2006 and fall of 2016**
*Source*: database of *Yahoo Finance*. URL: https://finance.yahoo.com/quote/%5EDJI/history?p=%5EDJI.

On the other hand, in a number of seminal works by Didier Sornette, Anders Johansen and their colleagues (Sornette, Sammis 1995; Sornette, Johansen 1997, 2001; Johansen, Sornette 1999, 2001; Johansen *et al.* 1996; Sornette 2004; etc.) it has been demonstrated that what is known as "stock exchange bubbles" can be quite well described mathematically. They have shown that accelerating log-periodic oscillations superimposed over an explosive growth trend that is described with a power-law function with a singularity (or quasi-singularity) in a finite moment of time $t_c$, are observed in situations leading to crashes and catastrophes. They can be analyzed because their precursors allow the forecasting of such events. One can mention such examples as the log-periodic oscillations of the Dow Jones Industrial Average (DJIA) that preceded the crash of 1929 (*e.g.*, Sornette, Johansen 1997), or the changes in the ion concentrations in the underground waters that preceded the catastrophic Kobe earthquake in Japan on the 17th of January, 1995 (*e.g.*, Johansen *et al.* 1996), which are also described mathematically rather well with log-periodic fluctuations superimposed over a power-law growth trend.

The basic equation derived by Sornette and tested on many historical examples of bubbles has the following form:

$$\ln[p(t)] = A - m\,(t_c - t)^\alpha \{\,1 + C\cos[\omega\ln(t_c - t) + \varphi]\,\}, \qquad (1)$$
or
$$p(t) = A - m\,(t_c - t)^\alpha \{\,1 + C\cos[\omega\ln(t_c - t) + \varphi]\,\}, \qquad (1a)$$

where $p(t)$ is the value of a certain financial indicator at the moment $t$; $t_c$ is the "critical time"; $A$, $m$, $C$, $\alpha$, $\omega$, and $\varphi$ are constants which are to be defined on the basis of data on gold prices from the start of the bubble formation till the forecast moment.

In these equations $m\,(t_c - t)^\alpha$ describes the main trend of the growth dynamics; with the approaching of the critical point $t_c$, price $p(t)$ approaches the maximum value $A$. Against this background periodic oscillations with reduced period take place. These oscillations (Sornette calls them log-periodic oscillations) are described by the second member $C\,(t_c - t)^\alpha \cos[\omega\ln(t_c - t) + \varphi]$ multiplied by the first considered member (whose value decreases with the lapse of time). Thus, the oscillations amplitude is steadily declining.

Of course, not only infinite, but also very high frequency of the financial market price fluctuations is really impossible. Achieving the oscillation frequency of high values means an increased risk of crisis. On average, in the examples treated by Sornette (2004), the crisis occurs 1.4 months before the critical point $t_c$.

At the moment of crisis, the growth of $p(t)$ stops and there frequently begins its sharp decline (market crash). The second option marked by Sornette is a scenario when the bubble softly "blows off", the price starts to decline more or less in reverse order with respect to the sequence in which it grew up ("anti-bubble"). This scenario is the least damaging to the economic and socio-political perspective and in order to direct the process to a smoother path various state regulatory mechanisms are used.

In the practical implementation of the described methods of forecasting there rises the question of choosing the time interval for which the parameterization will be implemented. In Sornette's monograph (2004) this issue is solved empirically, *i.e.* such an interval is chosen within which both the main growth trend and the oscillations with reduced periods described by equation (1) are clearly visible.

We will use this method to calculate the date of a Juglar downturn in American economy and world economy as a whole. Below one might see the results of time series approximations, starting from 2009 (when the new bubble was formed) (see Fig. 8). (In calculations equation (1a) has been used).

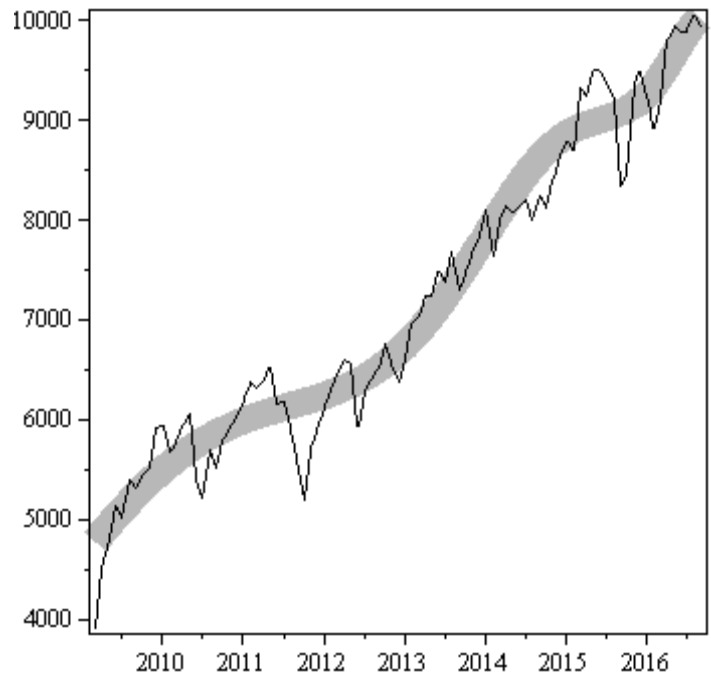

**Picture 8. Log-periodic oscillations in dynamics of DJIA between spring of 2009 and spring of 2016**

*Note*. The thin black curve corresponds to empirical data. The thick grey curve has been generated by the following equation with parameter values, defined through the least squares method: $p(t) = 10890.6 - 854.392 \cdot (2017.80 - t)^{0.950} + 85.600 \cdot (2017.80 - t)^{0.950} \cdot \cos[14.928 \cdot \ln(2017.80 - t) + 0.641]$. Calculations by Alexey Fomin.

Here the singularity point equals 2017.80 (= 10.19.2017). Thus, in accordance with these calculations, the nearest Juglar crisis in the American economy is likely to start in the fall of 2017.

It is noteworthy, that calculations related to a longer period (starting from the end of the Great Depression in the early 1930-s) provide us with a different singularity point at approximately 2046 (see Fig. 9). This point helps forecast a potential ending of the upswing phase of the sixth K-wave.

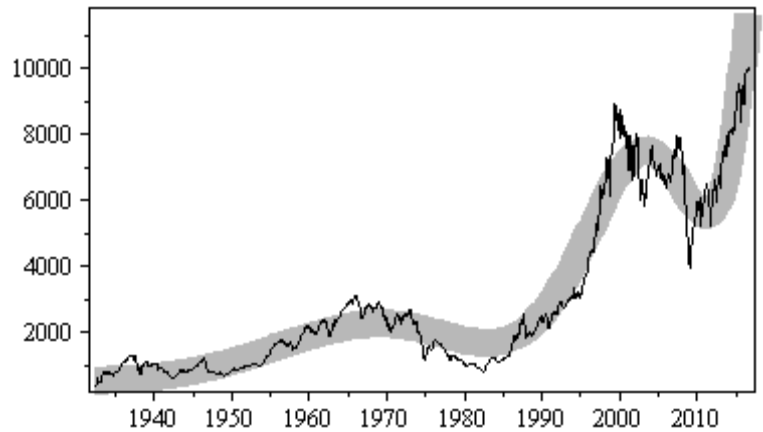

**Fig. 9. Log-periodic oscillations in dynamics of DJIA between spring of 1933 and spring of 2016**

*Note*. The thin black curve corresponds to empirical data. The thick grey curve has been generated by the following equation with parameter values, defined through the least squares method: $p(t) = 31.214 + 1.22 \cdot 10^7 / (2045.853 - t)^{2.047} + 4.74 \cdot 10^6 / (2045.853 - t)^{2.047} \cdot \cos[24.202 \cdot \ln(2045.853 - t) + 2.341]$. Calculations by Alexey Fomin.

**Beginning of the expansion phase of the sixth K-wave. Forecast**

Japanese engineer and economist Hirooka pointed out, that current economic and technological paradigm consists of three trajectories (see Fig. 10), technology, development and diffusion [*Hirooka, 2006*]. Technology encompasses the key technologies, related to a certain discovery, which came about as a result of technical or scientific breakthrough. Development trajectory consists of new innovative products, developed with the use of key technologies. Development proves to be the most crucial trajectory, as knowledge is transmitted from R&D institutions to industry, and new ventures are established to produce and eventually commercialize the new product. Favorable conditions for ventures usually take place within the first 10-15 years of the first half of the trajectory of development.

During this period, straight after the ending of the technology trajectory, comes about an intensive diffusion of an innovation into the market, which usually lasts for 25-30 years up to the market saturation.

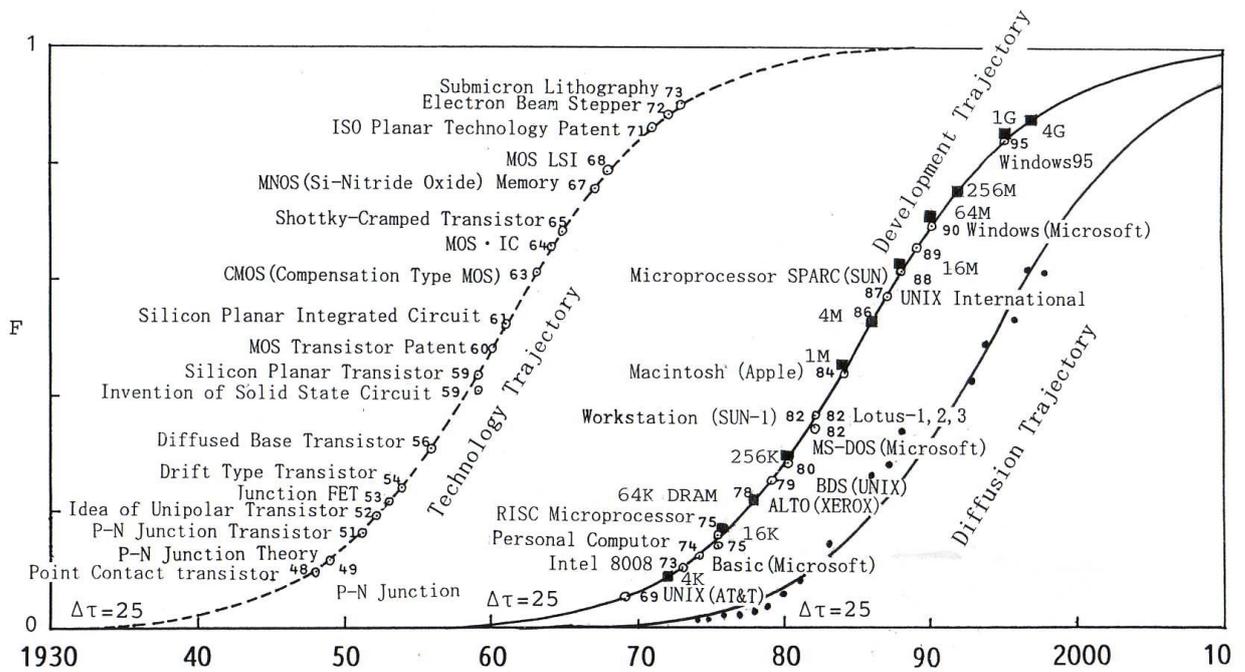

**Fig. 10. Electronics innovation paradigm**

Hirooka singled out and examined trajectory of development and showed it is represented with a logistic curve and lasts for approximately 30 years, beginning with a technological or scientific discovery. Hence, an innovation paradigm has a cascade structure, which consists of three logistic trajectories that have certain estimated distance between each other. This feature of an innovation paradigm enables more or less precise forecasting of diffusion of innovations into the market in accordance with previously defined technology trajectory (see Fig. 10), as is shown using the example with electronics [*Hirooka, 2006*]. The latter is 25-30 years ahead of the former, so it can be estimated before innovations enter the market. It is clear from Fig. 10, that technology of microelectronics began in 1948 when John Bardeen, Walter Brattain, and William Shockley introduced a transistor, a semiconductor device, and ended in 25 years, in 1973 when IBM introduced submicron lithography technique, which led to production of microchips, consisting of hundreds of millions of transistors. In 1973 Intel introduced the first microprocessor 4004, which contained 2300 transistors in silicon crystal. In 1974 the upgraded microprocessor with operation memory of 256 bytes Intel-8008 was introduced. It could run more than 75 commands. That microprocessor was used to develop the first personal computer in 1974. In two years, in 1976 Wozniak and Jobs introduced *Apple,* the first personal computer. Its mass use spread between 1977 and 1978. Thus, it is clear that within 25-30 years after the introduction of a new basic technology, come about new innovative products, which form new markets.

Using Hirooka's innovation paradigm, we will try to forecast the beginning of the expansion phase of the sixth K-wave. We know precisely that the development trajectory of nanotechnologies began in 1985, when fullerene was introduced and synthesized. Fullerene is a molecule of carbon $C_{60}$, consisting of 60 carbon atoms, which form a three-dimensional frame. It is the first artificial nanostructure. An atomic-force microscope was introduced in 1986. Its high resolution let see separate atoms and manipulate them. The microscope became the main instrument for creation of new nanostructures and their dimensions [*Williams, Adams, 2007*]. Later, the leading countries conducted further research in this sphere, which resulted in major inventions. Thus, as for the nanostructures, the following discoveries took place: in 1991 carbon nanotubes were introduced and later implemented in various spheres; in 2004 was discovered graphene, a flat carbon lattice one atom thick, which is now used in nanoelectronics. Due to their unique qualities, nanomaterials or "smart materials" are used in almost all the spheres of human activities, contributing to breakthroughs [*Rudskoy, 2007*].

Nanoinstruments were constantly being upgraded. Thus, came about scanning probe microscopes (SPM) with computer control, which allowed manipulating nanoparticles; optical tweezers for acquiring moving nanostructures in three-dimensional space; nanomanipulators with piezo motors, allowing smooth and controlled movement in any direction. In other words, at present nanoinstruments have reached an extremely high level of perfection, which paves the way for introduction of new nanostructures with changed characteristics and new practices. Nanoinstruments for industrial use were also introduced to produce nanoparticles and nanomaterials according to existing demand [*Williams, Adams, 2007*].

Thus, we have ascertained, that nanotechnologies (nanomaterials and nanoinstruments) are developing successfully and in accordance with Hirooka's paradigm (see Fig. 10), which probably reached saturation point in 2016, about 30 years after the SPM was introduced in 1986. Hence, according to Hirooka's paradigm, the technology trajectory was finished in 2016, and in upcoming years one expect the start of a large-scale diffusion of innovative nanoproducts into markets, which will eventually lead to economic growth in developed nations, and later the global economic growth. In the work [*Akaev, Pantin, Ayvasov, 2009*], based on innovation and cyclic theory of economic development of Schumpeter and Kondratieff, it was shown, that current depression will be protracted and last up to 2017-2018, when the expansion phase of the sixth K-wave will start.

Hence, it is possible to state, that *economies of the developed nations have experienced a certain revival since 2014-2015. Starting from 2018-2019, after the crisis of 2017 one could expect the start of the upswing phase of the sixth K-wave, defined by strong influence of the sixth techno-economic paradigm with its core of NBIC-convergence technologies. Therefore, governments of developed nations being the key actors in this sphere would rather concentrate all the resources and efforts to master NBIC-technologies, which form the sixth wave of innovation and the entire structure of the world economy. Period from 2017 to 2020-2024 is the most favorable time to master and spread new innovations on the basis of NBIC-technologies* [*Akaev, Rudskoy, 2013*].

NBIC-technologies develop with annual growth of 24%. According to the National Science Foundation, in 2015 the world market of nanoproducts and nanotechnologies totaled over 1 trillion dollars. Most was contributed by nanomaterials (31%), nanoelectronics (28%) and pharmaceuticals (17%). A new branch of the world economy with a high R&D intensity and more than 2 million highly qualified jobs was formed. We might expect an impressive start of new branches on the basis of NBIC-technologies, which will become the locomotive of the sixth K-wave.